# Local Determination of the Light deflection in the Spherically Symmetric Static Gravitational Field

**S.I. Tertychniy** *

Institute of Theoretical Physics
University of Cologne
50923 Cologne
Germany

**ABSTRACT**
The new method of invariant definition of the measurable angle of light deflection in the static central symmetric gravitational field is suggested. The predicted pure gravitational contribution to the deflection angle slightly differs from its classical estimate and one may hope that this discrepancy could be experimentally detected in the near future.

## 1. Introduction

The phenomenon of light deflection in a weak static central symmetric gravitational field is one of the most simple effect predicted by the general relativity that can be really checked in experiment. The first calculation of the gravitational light deflection in frames of general relativity has been performed by A.Einstein (ref. [1], Eq. (74)), and the recent version of his formula more appropriate for description of experiment has been obtained in [2] and [3]. Up to now the theoretical prediction and experimental results have not revealed any noticeable disagreement. Probably this is a reason why the theoretical calculations which are expounded in a number of textbooks and lead to the common wellknown result are likely never be undergone sufficiently careful analysis concerning the connection between the geometrical tools of the theory and the results of the physical measurements. Otherwise it is difficult to explain why it has not been mentioned in the literature that the main formula determining the *experimentally measurable* value of the gravitational light deflection must be revised. Nowaday the accuracy of experiment [4] has almost achieved however the threshold (about $10^{-8}$ radians, see Sec. 3 and 5 below) of a detectability of the distinction between the classical prediction and more consistent one obtained in this paper (Eqs. (22) and (A22) below).

Let us be more specific however. In the previous work by the author [5] the method of the invariant geometrical description of the deflection of light rays in the gravitational field of a static central symmetric massive body has been developed which, essentially,

* On leave of absence from: The National Research Institute For Physical Technical and Radio-Technical Measurements (VNIIFTRI), Mendeleevo, Moscow Region, 141570, Russia

operates with the quantities measurable in principle. To that end, it has been suggested to introduce the notion of the physical Riemannian space that, under the restrictions assumed, exhaustively characterizes the basic *space* relations available to observers and referring to the notion of *a distance* (including the angle between the space directions in particular). If one adopts such an approach then the sense of the light ray *deflection* becomes quite clear. Indeed, the shortest and at the same time straightest line in the physical space is its geodesic (the $p$-geodesic for brief) and the deflection of the light ray (*i.e.* the projection of the null space-time geodesic, the $s$-geodesic for brief, into the physical space) means precisely its deviation from the $p$-geodesic of the same initial direction. Such a deviation really occurs in a generic situation and one may suggest some quantitative characteristics describing it including, in particular, the total deflection angle. This approach seems to be the most natural *global geometric* interpretation of the effect of a distortion of a light beam by the static gravitational field. There are some reasons to name the corresponding deflection definition *global* (see also [5] for more details).

In practice, however, the principle of the measurement of the light deflection angle is absolutely different as well as the implicit definition of the deflection angle itself. It is in no way surprising since the sufficiently long pieces of a light beam trajectories are not really available to observer and it is difficult 'to stretch' the $p$-geodesics in the space in order to compare them with the light beams trajectories. Besides, the observations of the disposition of *the images* of removed sources *on the celestial sphere* of an individual observer are used. The deflection angle is realized as the arc between the 'usual' (or 'true') position of the image of the removed source on the celestial sphere and its position when the limb of the central massive body (the Sun as a rule) is close to it. Such a definition of the deflection angle can be named *local* in contrast to mentioned above since, in particular, the observer need not leave the local observatory where he observes the sky.

The latter remark does not mean however that the concept of the physical space cannot be applied to the description of the light deflection measured by the local method and we shall consider here the corresponding theoretical background. Besides the 'pure gravitational' deflection, the influence of the orbital motion of the observer is taken into account in a uniform way and this generalization does not lead to any noticeable complications.

We shall find that the classical formula determining the deflection angle which can be found in many wellknown textbooks (see, for example, [6-11] and also the discussion in the Section 5 below) *should be modified* even if there were *no* rotation of observer around the field center.

It is perhaps worthwhile to emphasize here that the new result is derived just in the frames of the standard general relativity and is merely due to the more consistent and essentially invariant treatment of the connection between geometrical tools of the theory and the results of physical measurements. The magnitude of discrepancy between the classical and new predicted values is expected to be on a level of the current experiment uncertainty and thus one may hope that it could be experimentally distinguished in the near future.

The parer is organized as follows. In the Section 2 we obtain the basic system of equations ensuring the rigorous definition of the locally measured light deflection angle. The appropriate Taylor expansion yielding the explicit formula is considered in the Section

3. It is fitted to the case of the Sun-Earth system in the Section 4. In the Section 5 we discuss the interpretations and the order of magnitudes of the separate terms in the resulting formula. In the Appendix the possible alternative arrangement of the measurement of deflection angle is briefly analyzed.

## 2. The physical metric and positions of light sources on the celestial sphere

Let us consider a static spherically symmetric space-time whose metric, when restricted to the hyperplane $z = 0$, has the form

$$\boldsymbol{ds}^2 = F(r)dt^2 - G(r)dr^2 - H(r)d\phi^2. \tag{1}$$

We drop the $z$-dimension since it will be sufficient for us to consider the motions of observers and the light rays geometry in the hyperplane $z = 0$.

We assume the space-time to be asymptotically flat so if one chooses $r$ coordinate in such a way that $H = r^2$ then $F$ and $G$ will tend to 1 as $r$ tends to infinity.

Further, let some space-time area be filled up continuously by *a family of observers* uniformly rotating around the center $r = 0$ of the field in accordance with the equations

$$\phi = \omega t + q, \quad r = P(p). \tag{2}$$

Here $\omega$ is some constant, $p$ and $q$ are the parameters uniquely identifying every individual observer (the usual identification *mod* $2\pi$ for the angle parameter $q$ is assumed). The function $P(p)$ is to be determined from the equation

$$p = \int^{P(p)} \sqrt{G(r)}dr.$$

A choice of the integration constant is clearly immaterial here.

The exchange of light signals by every pair of neighboring observers labeled by the parameters $(p, q)$ and $(p + dp, q + dq)$ respectively shows that the distance $\boldsymbol{dl}$ between them does *not* depend on a moment of measurement and is determined by the quadratic form

$$\boldsymbol{dl}^2 = dp^2 + \frac{H(r)F(r)}{F(r) - \omega^2 H(r)} dq^2, \tag{3}$$

where $r$ should be considered as the function of $p$. We use the system of units where the speed of light is equal to 1 and assume $F(r) - \omega^2 H(r) > 0$ in the area under consideration (in the flat space the latter would merely mean that the orbital speed of observers does not exceed the speed of light). The worldlines (2) of observers are obviously the orbits of the Killing vector field that automatically implies the constancy of the distance (3) 'in the course of time'.

We see therefore that the observers sense themselves to be immersed in *the static Riemannian space* with the metric (3). It is obviously the only representation of the ground level properties of metricity of *the space* available to them, the latter statement being equivalent to the claim that the speed of light measured in local experiments is the world constant even if the space-time is curved.

We shall name the Riemannian space endowed with the chart $(p, q)$ and the metric (3) the *physical space* and the *physical metric* respectively.

Now we remind that the null geodesic of the metric (1) is governed by the equations

$$d\phi = \epsilon\lambda \left[GF \left(1 - \lambda^2 F/H\right)^{-1}\right]^{1/2} H^{-1} dr,$$
$$dt = \epsilon \left[GF^{-1} \left(1 - \lambda^2 F/H\right)^{-1}\right]^{1/2} dr,$$

where the positive branch of square root is assumed. $\lambda$ is the constant which is usually called the impact parameter of the ray. The sign of $\lambda$ determines whether the coordinate $\phi$ locally increases or decreases along the (oriented) trajectory of photons. Symbol $\epsilon = -$ for the initial half of the ray when the photons approach the center and $\epsilon = +$ for the final half of ray when it tends to infinity. (We consider only the rays that are not seized by the gravitational field in the central region.)

In accordance with the Eqs. (2) the rotating observers find the following relation governing the trajectory of the light rays (projections of oriented null space-time geodesics)

$$\Omega(\boldsymbol{v}) = 0, \qquad \text{where} \quad \Omega \equiv dq - \epsilon \left[\lambda \left(F/H\right)^{1/2} - \omega \left(H/F\right)^{1/2}\right] \left(H - \lambda^2 F\right)^{-1/2} dp,$$

and $\boldsymbol{v}$ is the physical space vector tangent to the ray.

We see that the *direction* of the light ray is described for rotating observers by the following vector of *the physical space* which is tangent to the ray, normalized to the unit and annihilates the form $\Omega$:

$$\boldsymbol{v} = \boldsymbol{V}_\omega(\lambda, \epsilon) = \epsilon T(\lambda, \omega)\boldsymbol{X}_1 + W(\lambda, \omega)\boldsymbol{X}_2,$$

where

$$\begin{aligned} T(\lambda, \omega) &= (1 - \lambda\omega)^{-1}(HF)^{-1/2} \left[\left(F - \omega^2 H\right)\left(H - \lambda^2 F\right)\right]^{1/2}, \\ W(\lambda, \omega) &= (1 - \lambda\omega)^{-1} \left[\lambda \left(F/H\right)^{1/2} - \omega \left(H/F\right)^{1/2}\right], \\ \boldsymbol{X}_1 &= \partial_p, \\ \boldsymbol{X}_2 &= (F - \omega^2 H)^{1/2}(FH)^{-1/2}\partial_q. \end{aligned} \tag{4}$$

The pair of vectors $\{\boldsymbol{X}_1, \boldsymbol{X}_2\}$ is chosen to constitute the orthonormal frame in the physical space.

The vector $\boldsymbol{v}$ is precisely the *vectorial light speed* as it is perceived by the rotating observer and whose direction coincides with the 'mechanical' one of the axis of a telescope directed to the source. The observed angle between the images of two sources on the celestial sphere is nothing else but the angle between the corresponding light speed vectors in the point of observation. Further, the angle between the vectors tangent to two light rays having the common point and attached to that point has to be determined with respect to the physical metric and the cosine of the angle equals their scalar product. Thus if the rays are characterized by the parameters $(\epsilon', \lambda')$ and $(\epsilon'', \lambda'')$ respectively the angle $'\nu''$ between them is determined by the equation

$$\exp(i\,'\nu'') = [\epsilon' T(\lambda', \omega) + iW(\lambda', \omega)][\epsilon'' T(\lambda'', \omega) - iW(\lambda'', \omega)]. \tag{5}$$

This equation is the basic one yielding in particular the unique *rigorous invariant* interpretation of the notion of angle between two removed light sources as it is perceived by the chosen family of observers. Its sense becomes more transparent if one uses the dualization operator $*$ defined in action on the orthonormal frame as follows: $*\boldsymbol{X}_1 = \boldsymbol{X}_2$, $*\boldsymbol{X}_2 = -\boldsymbol{X}_1$; then $\exp(i\,'\nu'') =< \boldsymbol{v}_1|\boldsymbol{v}_2 > + \, i < \boldsymbol{v}_1| *\boldsymbol{v}_2 >$ ($< | >$ denotes the scalar product in the metric (3)).

Now we assume that the space-time is vacuum outside some region around the center occupied by the static massive body, *i.e.* $F = G^{-1} = 1 - 2m/r$, $H = r^2$ for $r$ greater than some $r_{min} > 3m$. The most of the reasoning below remains however valid in the case of arbitrary asymptotically flat space-time as well and one will easily distinguish the generic assertions.

Let us consider the following observation scheme (see the Fig. 1). The observer which determines the positions of removed radiating sources (stars or quasars) on his celestial sphere uniformly rotates around the central mass along the orbit with some constant value of the coordinate $r > r_{min}$. He can be included in a natural way into a family of uniformly and coherently rotating observers. We assume that the distances and angles are measured by observers in the manner described above. This is adequate in the frames of relativistic theory (up to the inevitable idealizing) to the corresponding practical means.

Further, we assume for simplicity that there exist two light sources exactly in the plane of observer's orbit with the arc separation between them to be not very small. They are periodically eclipsed by the central massive body. We shall call one of them the *master* source and another the *reference* source for brief.

One has the following natural **definition** of the light deflection angle. *It equals the difference of the arc separations between the images of the master and reference sources measured in two cases, first, when the master source is close to the limb of the central body while the reference source is comparatively far from it and, second, when the both sources are removed from the limb (on the celestial sphere) as far as possible.*

Thus the observer needs carry out two observations. The first of them is timed to the moment close to the eclipse of the master source by the limb of the massive central body. We shall call it the *principal* observation **P**.

The second, *calibrating* observation **C** is carried out in a typical situation approximately a half of a revolution period after (or before) the principal one. Then the parts of the both null geodesics forming the images of the master and calibrating sources lie completely outside the observer's orbit and are removed *in the space* as far as possible from the center body. At the same time the images of the master and reference sources are maximally removed from the limb of central body *on the celestial sphere.*

Let the observer register during the principal observation the following four light rays.

The first is the ray $\mathcal{L}$ from the master source passing near the central body and let it be characterized by the (unknown) $\lambda$-parameter $\lambda_0 > 0$. It is obvious that the ray $\mathcal{L}$ has passed the point closest to the center when it is detected by observer (*i.e.* $\epsilon = +$).

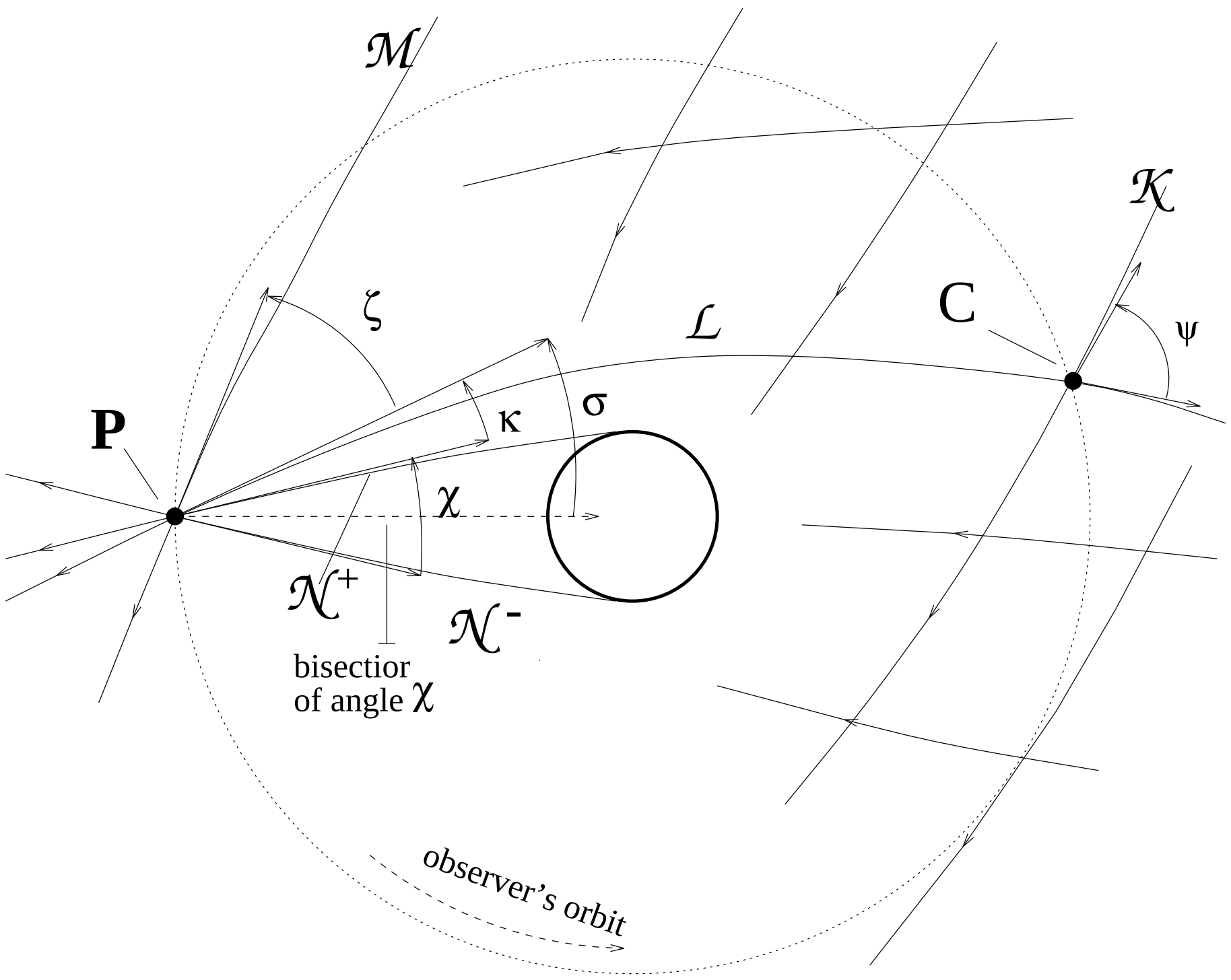

Figure 1. The sketchy picture of the relevant light rays in the plane of the observer's motion is presented. The crossed inclined and horizontal lines denote the congruences of rays emanated by the calibrating and master sources respectively. Dotted circular line is the orbit of observer, the less circle around the center represents the section of surface of the central massive body. Other notations are explained in the text. The arrows attached to the points **P** and **C** represent the tangent vectors to the rays but it is important to remember that the depicted angles between them (denoted by directed arcs) have no physical meaning and *do not coincide* with measurable angles that, in particular, depend on $\omega$.

The second one is the ray $\mathcal{M}$ from the reference source detected at the same moment of the principal observation. Let it correspond to the $\lambda$-parameter $\lambda_2 > 0$. We assume for definiteness that the ray $\mathcal{M}$ has also passed the point closest to the center (although the opposite assumption is also admissible). This means mainly that the angle $\zeta$ between the rays $\mathcal{L}$ and $\mathcal{M}$, coinciding with the arc separation between the master source and the reference source on the celestial sphere during the principal observation, is somewhat less than the right angle at least. This restriction is in fact technical in nature but it covers a wide class of the physically meaningful conditions of measurements.

We assume additionally that the angle $\zeta$ is not very small since otherwise the both disturbed and undisturbed angles between the sources images clearly vanish together with their difference.

The third and the fourth rays $\mathcal{N}^+$ and $\mathcal{N}^-$ registered by the observer are emanated by the points of intersection of the observer's orbit plane and the boundary of the limb of the central body. The $\lambda$-parameters $\lambda_3 > 0$ of the ray $\mathcal{N}^+$ and $-\lambda_3 < 0$ of $\mathcal{N}^-$ differs in sign ($\lambda_3$ is also unknown). The angle $\chi$ between $\mathcal{N}^+$ and $\mathcal{N}^-$ yields the arc diameter of the limb. Another relevant measured angle is the arc separation $\kappa$ between the image of the master source and the point of the limb closest to it,*i.e.* the angle between the rays $\mathcal{L}$ and $\mathcal{N}^+$. In fact, we shall not need know the angles $\kappa$ and $\chi$ separately. As we shall see below, it is the angle $\sigma \equiv \kappa + \chi/2$ between the master source and the center of the limb which really enter the relevant equations and crucially determines the purely gravitational deflection angle.

All the $\epsilon$-parameters of the rays $\mathcal{L}, \mathcal{M}, \mathcal{N}^+, \mathcal{N}^-$ equals $+$ (*i.e.* the $r$ coordinate increases along all the rays near the point $\mathbf{P}$).

The arc separation $\psi$ between the master and reference sources registered during the *calibrating* observation may be considered to be undisturbed or, to be more exact, minimally disturbed by the gravitational field comparatively to the other possible choices of the instant of observation. Hence the discrepancy $\delta \equiv \psi - \zeta$ may be interpreted to be caused just by the gravitation deflection of the rays from the master source when they are passing through the region nearby the central body (and the influence of the observer's orbital motion as well).

The angle $\delta$ is constructed essentially from the values directly measured in experiment and it seems to be the most natural candidate to the role of the rigorous notion of the observable gravitational deflection angle. This definition of the deflection angle will be named *local* since all the measurements are carried out over the arbitrarily small parts of the light rays (formally, over the tangent vectors to rays).

We assume for simplicity that the projections of geodesics forming the images of the master source are *the same* for the both observations, *i.e.* it is the ray $\mathcal{L}$ from the master source which is registered during the calibrating observation. In particular, it corresponds to the same $\lambda$-parameter $\lambda_0 > 0$ (meanwhile the $\epsilon$-parameter is opposite: $\epsilon = -$).

Our assumption fixes the instant $\mathbf{C}$ of the calibrating observation unambiguously. It seems also to be plausible from a geometrical point of view. Indeed, one may say that it enables one to compare the directions of the motion of a photon just *before* and *after* it has been undergone the action of the gravitational field in the central region (see Fig. 1), the 'standard' for comparison being the congruences of rays from the reference source.

It seems worthwhile however to make the additional remark here.

There is still some uncertainty in the understanding of what a position of the observer corresponds to the minimally disturbed position of the master and reference sources on the celestial sphere (certain distortion is of course inevitable provided the gravitational field exist; to be more exact, it is impossible to define what a position of the source might be understood as 'undisturbed' one and our occasional use of this word is not, strictly speaking, quite legal). In some extent this is a matter of convention in fact. This makes our above definition of the deflection angle not absolutely certain. Intuitively, our choice allows to describe the deflection of a *single* ray which is registered during the principal observation. Its precise formulation given above is rather convenient for calculations and has more or less clear interpretation but perhaps significantly less suits the realization in practice. Moreover, it has a disadvantage to become somewhat obscure when $\sigma$ approaches the right angle and fails when $\sigma$ exceeds it.

On the other hand one may argue, especially for comparatively large $\psi$, that the positions of sources would be less disturbed provided the 'radial vector' * during the calibrating observation is the *bisector* of the angle between directions pointing the sources (see the Fig. 2). The 'bisectorial' choice of the instant of the calibrating measurement possesses also a pleasant feature to yield a *maximum* to the pure gravitational contribution (see Section 4 below) to the deflection angle, as a simple symmetry–based speculation shows.

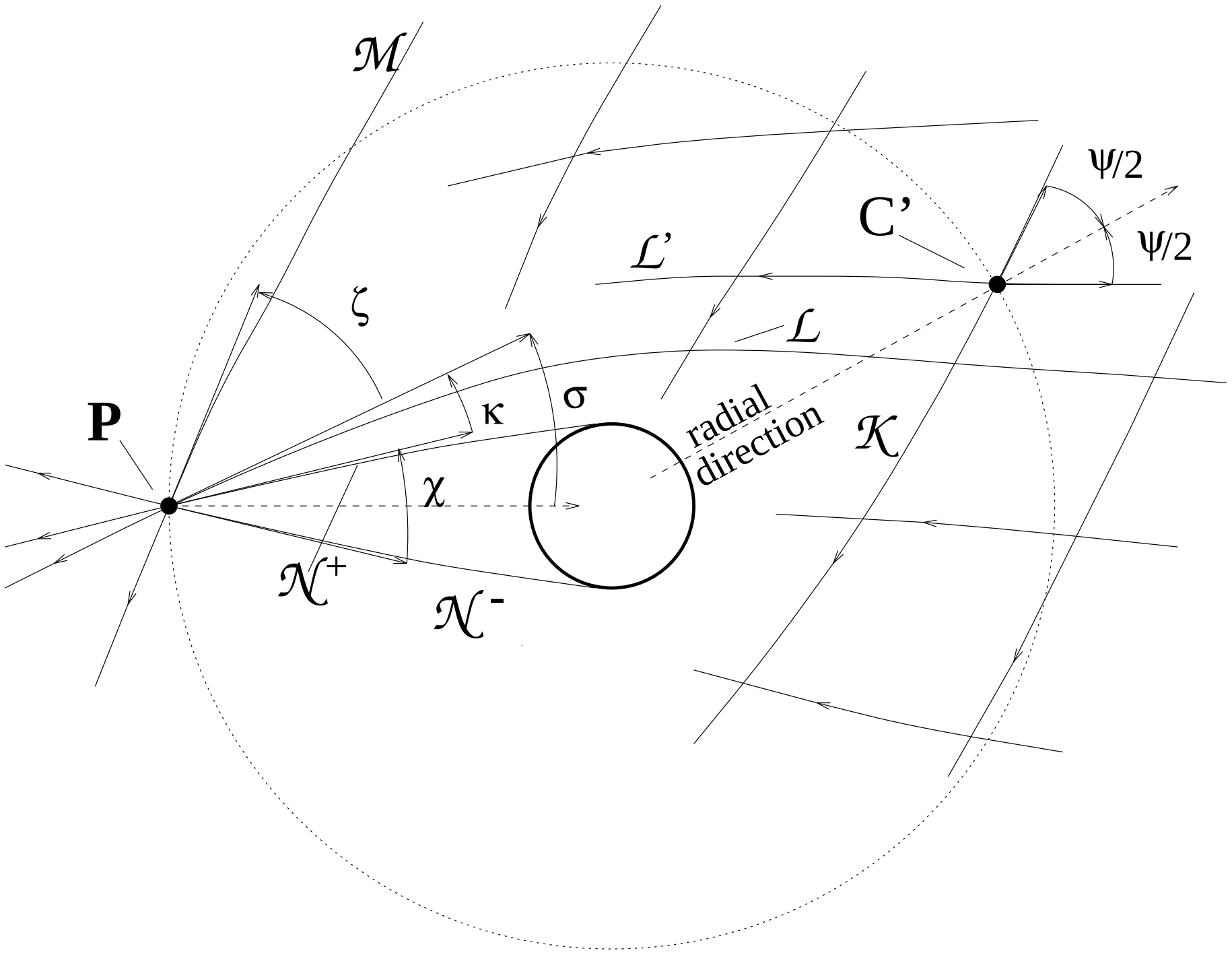

Figure 2. It is similar to Fig. 1 but the point **C'**, that replaces **C** of Fig. 1 and corresponds to the instant of calibrating observation, is chosen in such a way that the radial direction divides the angle constituting in **C'** by the rays from the reference and master sources into two equal parts.

Nevertheless we prefer here to make the definite choice mentioned above remembering that this point may require additional discussion. Besides, the 'bisectorial' choice of the moment of calibrating measurement is briefly analyzed in the Appendix where the corresponding version of the main resulting formula is derived.

Now let us continue. It is clear from the Fig. 1 that the rays $\mathcal{K}$ and $\mathcal{M}$ are 'almost parallel' and thus the $\lambda$-parameter of the ray $\mathcal{K}$ is negative provided the angle $\sigma + \zeta$ is not very small (we have assumed that $\zeta$ at least is not very small). We designate however $\lambda_1$

* The notion of the 'radial direction' (pointing to the center of central body) is not quite obvious in our case due to effect of relativistic aberration. It can be described in a constructive manner but will possess some unusual properties. To avoid undue complications, it is reasonable here to assume for a time that the rotation of observer is sufficiently slow to be neglected.

the absolute value of the corresponding $\lambda$-parameter and shall introduce the minus sign straight into the relevant formulae. $\epsilon = -$ for the ray $\mathcal{K}$ near **C**.

Thus as a result of the two observations the following four angles become known: the disturbed ($\zeta$) and undisturbed ($\psi$) arc distances between the sources, the arc diameter $\chi$ of the central body and the angle $\kappa$ between the master source and the limb edge during the principal measurement. Our goal is to express the *deflection angle* $\delta \equiv \psi - \zeta$ as a function of the parameters set in experiment $(\psi, \chi, \kappa, r, \omega)$ and the space-time characteristics (such as the mass $m$ of the central body). This can be achieved as follows.

In accordance with the Eq. (4) one has the following expressions for the vectors tangent to the light rays:

| | | | | |
|---|---|---|---|---|
| to the ray | $\mathcal{L}$ | in the point | **C** | ${}_C\boldsymbol{v}_0 = \boldsymbol{V}_\omega(\lambda_0, -)$, |
| | $\mathcal{K}$ | | **C** | ${}_C\boldsymbol{v}_1 = \boldsymbol{V}_\omega(-\lambda_1, -)$, |
| | $\mathcal{L}$ | | **P** | ${}_P\boldsymbol{v}_0 = \boldsymbol{V}_\omega(\lambda_0, +)$, |
| | $\mathcal{M}$ | | **P** | ${}_P\boldsymbol{v}_2 = \boldsymbol{V}_\omega(\lambda_2, +)$, |
| | $\mathcal{N}^+$ | | **P** | ${}_+\boldsymbol{v}_3 = \boldsymbol{V}_\omega(\lambda_3, +)$, |
| | $\mathcal{N}^-$ | | **P** | ${}_-\boldsymbol{v}_3 = \boldsymbol{V}_\omega(-\lambda_3, +)$. |

Then the Eq. (5) implies the following relations between the four basic angles defined above and light rays parameters:

$$\begin{aligned} \psi &= \arctan U(\lambda_1, -\omega) + \arctan U(\lambda_0, \omega), \\ \zeta &= \arctan U(\lambda_2, \omega) - \arctan U(\lambda_0, \omega), \\ \chi &= \arctan U(\lambda_3, \omega) + \arctan U(\lambda_3, -\omega), \\ \kappa &= \arctan U(\lambda_0, \omega) - \arctan U(\lambda_3, \omega). \end{aligned} \tag{6}$$

Here $\quad U(\lambda, \omega) \equiv W(\lambda, \omega)/T(\lambda, \omega)$

$$= (\lambda H^{-1/2} F - \omega H^{1/2})[(F - \omega^2 H)(1 - \lambda^2 F/H)]^{-1/2}.$$

The choice of the positive branch of the square root is assumed throughout.

Eqs. (6) establish the local 1-to-1 correspondence between the set of $\lambda$-parameters of rays and the set of measured angles.

The additional equation should express the fact of the intersection at a fixed $r$ value of the two rays $\mathcal{K}$ and $\mathcal{M}$ from the reference source parallel in asymptotic with the ray $\mathcal{L}$ from the master source. Indeed, one has the following dependencies of the coordinate $\phi$ on $r$ along the rays:

| | | | | |
|---|---|---|---|---|
| on the ray | $\mathcal{L}$ | near the point | **C** | $\phi = I(r, \lambda_0)$, |
| | $\mathcal{K}$ | | **C** | $\phi = \mathring{\phi} - I(r, \lambda_1)$, |
| | $\mathcal{L}$ | | **P** | $\phi = J(r, \lambda_0)$, |
| | $\mathcal{M}$ | | **P** | $\phi = \mathring{\phi} + J(r, \lambda_2)$. |

Here $\mathring{\phi}$ is an unknown constant (equal to the asymptotic angle between the master and calibrating congruences) and

$$
\begin{aligned}
I(r,\lambda) &= \lambda \int_r^\infty Z dr, \qquad Z = \left[GFH^{-1}\left(H-\lambda^2 F\right)^{-1}\right]^{1/2}, \\
J(r,\lambda) &= \lambda\Big(\int_{R(\lambda)}^\infty + \int_{R(\lambda)}^r\Big) Z dr \\
&= 2I(R(\lambda),\lambda) - I(r,\lambda).
\end{aligned}
\tag{7}
$$

$R(\lambda)$ denotes the maximal root of the equation $H = \lambda^2 F$; it is assumed to be simple.

The intersection of rays in **P** and **C** implies the equations

$$
I(r,\lambda_0) = \mathring{\phi} - I(r,\lambda_1), \qquad J(r,\lambda_0) = \mathring{\phi} + J(r,\lambda_2).
$$

Eliminating $\mathring{\phi}$, the equation follows

$$
\begin{aligned}
I(r,\lambda_0) + I(r,\lambda_1) = J(r,\lambda_0) - J(r,\lambda_2) \quad &\Leftrightarrow \\
2[I(R(\lambda_0),\lambda_0) - I(r,\lambda_0)] = I(r,\lambda_1) + 2I(R(\lambda_2),\lambda_2) &- I(r,\lambda_2).
\end{aligned}
\tag{8}
$$

Together with (6) one has the system of *five* equations. After the excluding of the *four* unknown $\lambda$-parameters, the relation between the angles $\zeta, \psi, \chi, \kappa$ can be established. If it is expressed in the form $\delta = \psi - \zeta = \delta(\psi, \chi, \kappa)$ the problem may be considered to be resolved.

It remains to reduce the solution to a form more suitable for applications in the typical situations.

## 3. The deflection formula

Let us notice that there are *two* small parameters in our formulae. The first of them is the ratio of the gravitational radius $2m$ of the central body to the $\lambda$-parameter minimal among $\lambda_j$ (if all the rays escape the central body then, in the case of Sun-Earth system, for example, one will have $m/\lambda_j \leq 10^{-6}$). Another small parameter is the 'speed' of the observer's rotation $(\omega r)$. In principle, it may be of any value less than the unit but in the case of a *free* motion along a circular orbit in the Schwarzschild field one has precisely $(\omega r)^2 = m/r$ (the generalized Kepler's law). Hence it is reasonable to assume that in any physically meaningful case $(\omega r)$ cannot be much greater than $\sqrt{m/r}$. At least, in the case of the Sun-Earth system one has as a typical value $\omega r \simeq 10^{-4}$.

Clearly, it is convenient to apply the Taylor expansion with respect to the mentioned small parameters. Some care must be taken however to decide what the orders of terms are to be kept.

To that end, let us notice that the recent light deflection observations for the Sun-Earth system by means of the VLBI method confirms the value $\gamma = 1$ of the space curvature parameter $\gamma$ of the PPN approximation [6,7] with the accuracy about 0.002 [4]. The deflection angle is proportional to $1 + \gamma$ and, thus, permitting ourselves somewhat free interpretation, we may assume that the deflection angle can be experimentally measured with the accuracy within 0.1 % of its maximal value (1.7 arc seconds or $10^{-5}$ radians), *i.e.* $10^{-8}$

radians (see also [12], pp. 5,155). One may neglect therefore any contributions significantly less than this *measurement uncertainty* $\epsilon \simeq 10^{-8}$ radians without prejudice to the result. Hence, the only small parameters powers that have to be kept are $\omega r, (\omega r)^2, m/\lambda_j \sim m/r$ and, conditionally, $(\omega r)(m/\lambda_j) \sim m\omega$. All the other terms will be dropped.

We shall say that this approximation is *of the order* 3/2 (since the last kept term is typically estimated as $m\omega \sim (m/r)^{3/2}$).

The following expansion of the order 3/2

$$\arctan U(\lambda, \pm\omega) \simeq \alpha \mp (\omega r)\cos\alpha - (m/r)\tan\alpha - \tfrac{1}{4}(\omega r)^2 \sin 2\alpha \mp (m\omega)\sec\alpha,$$

where $\sin\alpha = \lambda, /r \;\; 0 \le \alpha < \pi/2$, and the Eqs. (6) yield the representations of the angles $\psi, \zeta, \chi, \kappa$ through the auxiliary unknowns $\alpha_j \equiv \arcsin(\lambda_j/r)$. The difference of the expansions of the two first equations (6) yields

$$\begin{aligned}\delta = \psi - \zeta \simeq \;\; & \alpha_1 + 2\alpha_0 - \alpha_2 \\ & + (\omega r)\Delta^+ [\cos(\cdot)] + (m/r)\Delta^- [\tan(\cdot)] \\ & + \tfrac{1}{4}(\omega r)^2 \Delta^- [\sin 2(\cdot)] + (m/r)(r\omega)\Delta^+ [\sec(\cdot)] .\end{aligned} \tag{9}$$

Here the following functionals are introduced for convenience:

$$\Delta^{\pm}[f(\cdot)] \equiv f(\alpha_2) \pm f(\alpha_1) - 2f(\alpha_0)$$

for every function $f(\alpha)$. The Eq. (9) determines the desirable deflection angle but its r.h.s. contains the unknown $\alpha$'s.

We need now the additional equation (8) in order to exclude them. Let us fit it to the Schwarzschild case when $F = G^{-1} = 1 - 2m/r, \;\; H = r^2$. After the substitutions, the corresponding indefinite version of the integrals $I, \;\; J$ (7) takes the form

$$\lambda \int \left\{ r \left[ r^2 - \lambda^2 (1 - 2m/r) \right]^{1/2} \right\}^{-1} dr = \lambda \int \left[ r\,(r-\rho_1)\,(r-\rho_2)\,(r-\rho_3) \right]^{-1/2} dr,$$

where $\quad \rho_1 = \lambda(\Xi - \xi), \quad \rho_2 = 2\lambda\xi, \quad \rho_3 = -\lambda(\Xi + \xi)$

and $\quad \xi \equiv \xi(\lambda) = 3^{-1/2} \sin\left(3^{-1}\left(\arcsin\left(3^{3/2} m/\lambda\right)\right)\right), \quad \Xi \equiv \Xi(\xi) = (1 - 3\xi^2)^{1/2}. \quad (10)$

This expression determines precisely the first kind elliptic integral but it is more convenient for us to work on with its approximate representation for small $\xi$ in terms of elementary functions.

To that end, let us notice that if $0 \le \xi < 1/(2\sqrt{3})$ the maximal root $R(\lambda)$ of the equation $H = \lambda^2 F$ equals $\rho_1$ and is simple. Passing to the new integration variable $\nu = [r/\lambda - \Xi + \xi]^{1/2}$ one obtains from (7)

$$I(R(\lambda), \lambda) = I_0(\xi(\lambda)) \equiv \int_0^\infty \Omega(\nu, \xi) d\nu,$$

$$I(r,\lambda) = I_1(\xi(\lambda), r/\lambda) \equiv \int_L^\infty \Omega(\nu,\xi) d\nu,$$

where $\quad \Omega = \left[\left(\nu^2 + \Xi - \xi\right)\left(\nu^2 + \Xi - 3\xi\right)\left(\nu^2 + 2\Xi\right)\right]^{-1/2},$

$$L = L(r/\lambda, \xi) = (r/\lambda - \Xi + \xi)^{1/2}.$$

The latter representations have the advantage to be explicitly smooth in the vicinity of the point $\xi = 0$ and the Taylor expansion can be applied to them straight. Furthermore, it is sufficient to keep only linear terms in the expansion. Indeed, in accordance with the definition of $\rho_1$,

$$\lambda = R(\lambda)/[(1-3\xi^2)^{1/2} - \xi] > R(\lambda) \quad \Rightarrow \quad m/\lambda_j < m/R,$$

provided $0 < \xi < 0.5$. On the other hand $R(\lambda)$ is the minimal value of the coordinate $r$ along the light ray and it at least exceeds the radius of the central body provided the ray escapes it. [This seems to be too restrictive for the rays $\mathcal{K}$, $\mathcal{M}$ but one can infer from the Fig. 1 that it holds if $\zeta$ and $\psi$ are not comparatively small]. Hence we may assume $m/R$ to be sufficiently small ($m/R \simeq 10^{-6}$ for the Sun) to neglect all its powers except the first. This is true for all $m/\lambda_j$ (and $m/r$) as well. Such a practice of the linear approximating with respect to the gravitational radius $2m$ is in general use in the calculations of deflection angle during estimations of the (7)-like integrals.

Under these conditions, one has in the first approximation

$$\xi \simeq m/\lambda \ll 1$$

and

$$\begin{aligned}
I_0(\xi) &\simeq I_0(0) + \xi I_0'(0) \\
&= \int_0^\infty \left\{\left(\nu^2+1\right)\sqrt{\nu^2+2}\right\}^{-1} d\nu + 2\xi \int_0^\infty \left\{\left(\nu^2+1\right)^2\sqrt{\nu^2+2}\right\}^{-1} d\nu \\
&= \pi/4 + \xi \simeq \pi/4 + m/\lambda = \tfrac{1}{4}\pi + \frac{m}{r}\operatorname{cosec}\alpha, \\
I_1(\xi, r/\lambda) &\simeq I_1(0, r/\lambda) + \xi \partial_1 I_1(0, r/\lambda) \\
&= \int_{L_0}^\infty \left\{\left(\nu^2+1\right)\sqrt{\nu^2+2}\right\}^{-1} d\nu \\
&\quad + \xi \left[-(\lambda/2r)\left((r/\lambda)^2 - 1\right)^{-1/2} + 2\int_{L_0}^\infty \left\{\left(\nu^2+1\right)^2\sqrt{\nu^2+2}\right\}^{-1} d\nu\right] \\
&= \tfrac{1}{2}\alpha + \frac{m}{r}\operatorname{cosec}\alpha(1 - \tfrac{1}{2}\cos\alpha - \tfrac{1}{2}\sec\alpha)
\end{aligned}$$

where $L_0 = (r/\lambda - 1)^{1/2}$ and, as above, $\sin\alpha = \lambda/r,\ \ 0 \le \alpha < \pi/2$.

Then the first order approximation to the Eq. (8) takes the form

$$\begin{aligned}
2\alpha_0 + \alpha_1 - \alpha_2 &+ (2m/r)[\operatorname{cosec}\alpha_1 + \operatorname{cosec}\alpha_2] \\
&+ (m/r)[\operatorname{cotan}\alpha_2 - \operatorname{cotan}\alpha_1 - 2\operatorname{cotan}\alpha_0] \\
&+ (2m/r)[\operatorname{cosec}2\alpha_2 - \operatorname{cosec}2\alpha_1 - 2\operatorname{cosec}2\alpha_0] \simeq 0.
\end{aligned} \tag{11}$$

This approximation can be considered to be of order 3/2 as well since the neglected terms $O((m/\lambda)^2)$ are of the second order.

Eq. (11) together with Eq. (9) imply the following one:

$$\begin{aligned}\delta \simeq \quad & (\omega r)\Delta^+ [\cos(\cdot)] \\ & + (m/r)\big\{\Delta^- [\tan(\cdot)] - \Delta^- [\text{cotan }(\cdot)] \\ & \qquad - 2\Delta^- [\text{cosec } 2(\cdot)] - 2\Delta [\text{cosec }(\cdot)]\big\} \\ & + \tfrac{1}{4}(\omega r)^2 \Delta^- [\sin 2(\cdot)] \\ & + (m\omega)\Delta^+ [\sec(\cdot)] .\end{aligned} \tag{12}$$

Here we have introduced additionally the functional

$$\Delta[f(\cdot)] \equiv f(\alpha_2) + f(\alpha_1).$$

The following remark concerning the Eq. (12) is useful. Its r.h.s. still contains the unknowns $\alpha_j$ but now all such terms have the small coefficients (at least of the order 1/2). Hence it is sufficient to substitute the *first* order representation for $\alpha_j$ in order to obtain the $\delta$ value up to *the order* 3/2.

The corresponding first order version of the Eqs. (6) is

$$\begin{aligned} & \chi \simeq 2\alpha_3 - (2m/r)\tan\alpha_3 - \tfrac{1}{2}(\omega r)^2 \sin 2\alpha_3, \\ & \psi \simeq \alpha_1 + \alpha_0 + \omega r(\cos\alpha_1 - \cos\alpha_0) \\ & \qquad - (m/r)(\tan\alpha_1 + \tan\alpha_0) - \tfrac{1}{4}(\omega r)^2(\sin 2\alpha_1 + \sin 2\alpha_0), \\ & \kappa \simeq \alpha_0 - \alpha_3 - \omega r(\cos\alpha_0 - \cos\alpha_3) \\ & \qquad - (m/r)(\tan\alpha_0 - \tan\alpha_3) - \tfrac{1}{4}(\omega r)^2(\sin 2\alpha_0 - \sin 2\alpha_3), \\ \rightarrow \quad & \sigma = \kappa + \chi/2 \simeq \alpha_0 - \omega r(\cos\alpha_0 - \cos\alpha_3) \\ & \qquad - (m/r)\tan\alpha_0 - \tfrac{1}{4}(\omega r)^2 \sin 2\alpha_0. \end{aligned} \tag{13}$$

.

The Eqs. (11), (13) constitute a closed system with the following *first order* solution

$$\begin{aligned} & \left.\begin{matrix}\alpha_0 \\ \alpha_1 \\ \alpha_2\end{matrix}\right\} \simeq \left\{\begin{matrix}\sigma \\ \psi - \sigma \\ \psi + \sigma\end{matrix}\right\} + (\omega r)\mathbf{A}_j + (m/r)\mathbf{B}_j + (\omega r)^2\mathbf{C}_j, \qquad j = 0, 1, 2; \\ & \alpha_3 \quad \simeq \chi/2 + (m/r)\tan(\chi/2) + \tfrac{1}{4}(\omega r)^2 \sin\chi. \end{aligned} \tag{14}$$

Here

$$\begin{aligned} & \mathbf{A}_0 = \cos\sigma - \cos(\chi/2), \quad \mathbf{B}_0 = \tan\sigma, \\ & \mathbf{A}_1 = \cos\chi/2 - \cos(\psi - \sigma), \quad \mathbf{B}_1 = \tan(\psi - \sigma), \\ & \mathbf{A}_2 = 2\mathbf{A}_0 + \mathbf{A}_1 = 2\cos\sigma - \cos\chi/2 - \cos(\psi - \sigma), \\ & \mathbf{B}_2 = 2\tan\sigma + 2\tan(\psi - \sigma) \\ & \qquad + 2\mathring{\Delta}[\text{cosec}(\cdot)] + \mathring{\Delta}^-[\text{cotan}(\cdot)] + 2\mathring{\Delta}^-[\text{cosec}2(\cdot)], \end{aligned}$$

and the functionals

$$\mathring{\Delta}^-[f(\cdot)] \equiv f(\psi+\sigma) - f(\psi-\sigma) - 2f(\sigma),$$
$$\mathring{\Delta}[f(\cdot)] \quad \equiv f(\psi+\sigma) + f(\psi-\sigma)$$

have been introduced for convenience. (The explicit expressions of $\mathbf{C}_j$ will not be used in what follows and are omitted here.)

After a substitution for these representations of $\alpha_j$ into the Eq. (12) one obtains

$$\delta \simeq (\omega r)\mathbb{A} + (m/r)\mathbb{B} + (\omega r)^2\,\mathbb{C} + (m\omega)\mathbb{D}, \tag{15}$$

where
$$\begin{aligned}
\mathbb{A} &= \mathring{\Delta}^+[\cos(\cdot)],\\
\mathbb{B} &= \mathring{\Delta}^-[\tan(\cdot)] - \mathring{\Delta}^-[\mathrm{cotan}(\cdot)] - 2\mathring{\Delta}^-[\mathrm{cosec}2(\cdot)] - 2\mathring{\Delta}[\mathrm{cosec}(\cdot)],\\
\mathbb{C} &= \mathring{\Delta}^-[\sin 2(\cdot)] - \hat{\Delta}[\sin(\cdot),\mathbf{A}],\\
\mathbb{D} &= \mathring{\Delta}^+[\sec(\cdot)] - \hat{\Delta}^+[\sin(\cdot),\mathbf{B}]\\
&\quad + \hat{\Delta}^-[\sec^2(\cdot),\mathbf{A}] + \hat{\Delta}^-[\mathrm{cosec}^2(\cdot),\mathbf{A}]\\
&\quad + 4\hat{\Delta}^-[\cos 2(\cdot)\mathrm{cosec}^2 2(\cdot),\mathbf{A}] + 2\hat{\Delta}[\cos(\cdot)\mathrm{cosec}^2(\cdot),\mathbf{A}].
\end{aligned}$$

Besides the $\Delta$-functionals being defined above we have used here

$$\mathring{\Delta}^+[f(\cdot)] \quad \equiv f(\psi+\sigma) + f(\psi-\sigma) - 2f(\sigma),$$
$$\hat{\Delta}^\pm[f(\cdot),\mathbf{Y}] \equiv \mathbf{Y}_2 f(\psi+\sigma) \pm \mathbf{Y}_1 f(\psi-\sigma) - 2\mathbf{Y}_0 f(\sigma),$$
$$\hat{\Delta}[f(\cdot),\mathbf{Y}] \quad \equiv \mathbf{Y}_2 f(\psi+\sigma) + \mathbf{Y}_1 f(\psi-\sigma).$$

Eq. (15) is the almost final result. It remains only to recast it to a decent form. After some transformations one can obtain the following rather compact representations for the first three coefficients in (15):

$$\mathbb{A} = -2(1-\cos\psi)\cos\sigma, \tag{16}$$
$$\mathbb{B} = \quad 4\,\mathrm{cotan}\,\sigma\frac{\sin\psi}{\sin\psi+\sin\sigma}, \tag{17}$$
$$\mathbb{C} = -2(1-\cos\psi)(\sin\psi\cos^2\sigma - \sin\sigma\cos\sigma + \sin\sigma\cos\chi/2). \tag{18$a$}$$

The second formula turns out to be surprisingly simple comparatively with its initial form.

As to the fourth coefficient $\mathbb{D}$, it seems unlikely that it could be reduced to a more simple and at the same time useful form than as it stands. Fortunately. we need not such a representation in fact since only the unique leading term in the whole $\mathbb{D}$ expression is valuable in a typical situation as we shall see below.

## 4. The light deflection in the Sun-Earth system

It is useful to remind here the orders of main parameters for the typical case of the observation from the Earth's orbit of the light deflection near the Sun. One has in this case

$$2m \approx 2.5 \text{ km}, \qquad \omega \approx 10^{-7} \text{ sec}^{-1}, \qquad r \approx 1.5\,10^{8} \text{ km}, \qquad \chi \approx 16' \approx 0.005 \text{ rad}.$$

and our two basic small variables of expansions are of order

$$\omega r \simeq 30 \text{ km/sec} \simeq 1.\,10^{-4} \ \mathbf{c} \text{ (light speed)}, \qquad m/r \simeq 1.\,10^{-8}.$$

Hence

$$(\omega r)^2 \simeq m/r \simeq 1.\,10^{-8}, \qquad m\omega \simeq (\omega r)^3 \simeq 1.\,10^{-12}.$$

[The first equality is also implied by the free character of the orbital motion of the Earth]. Further, as we have mentioned above, the basic level of the admissible uncertainty in our formulae is $\epsilon \approx 10^{-8}$ (radians). Let us analyze the Eq. (15) conformably to that case.

Formulae (16),(17) need no further simplifications.

The following obvious estimate holds for the third term of (15):

$$0 < -(\omega r)^2\,\mathbb{C} < 4\epsilon.$$

which thus less then 4 times exceeds the threshold of detectability. In accordance with this estimate one may zero $\chi$ in the $\mathbb{C}$ expression ($18a$) for any admissible $\psi$. This may cause an error not more than $\epsilon\chi^2$ in magnitude that one may neglect. Thus we may assume

$$\mathbb{C} \approx -2(1-\cos\psi)[\cos^2\sigma\sin\psi + \sin\sigma(1-\cos\sigma)]. \qquad (18b)$$

Further, the most interesting is the case when the pure gravitational contribution to the angle $\delta$ is maximal. As we shall see below it occurs when $\sigma$ (restricted from below by $\chi/2 \approx 0.0025$) is minimal. In such a case of small $\sigma \sim \chi$ one may use a more rough approximation which is yielded by the setting $\sigma = 0$ in ($18b$). Then the simplest representation of $\mathbb{C}$ follows:

$$\mathbb{C} \approx -2(1-\cos\psi)\sin\psi. \qquad (18c)$$

Finally, let us revert to the Eq. (15) and consider its last term. At first, we must notice that the expression defining $\mathbb{D}$ contains the terms diverging when $\psi = \sigma$, the case quite admissible from the both physical and geometrical points of view. One can distinguish the following two apparently diverged aggregates involved in $\mathbb{D}$: the first can be extracted from the last three terms of the expression defining $\mathbb{D}$ and is equal to

$$-4\mathbf{A}_1\cos 2(\psi-\sigma)\text{cosec}^2 2(\psi-\sigma) - \mathbf{A}_1\text{cosec}^2(\psi-\sigma) + 2\mathbf{A}_1\cos(\psi-\sigma)\text{cosec}^2(\psi-\sigma),$$

and the second is the fragment of $\mathbf{B}_2$ equal to

$$-\ \text{cotan}(\psi-\sigma) - 2\text{cosec}\,2(\psi-\sigma) + 2\text{cosec}\,(\psi-\sigma).$$

They however can be recast to

$$\mathbf{A}_1(2\cos(\psi-\sigma)+1)\sin^2\tfrac{1}{2}(\psi-\sigma)[\cos(\psi-\sigma)\cos\tfrac{1}{2}(\psi-\sigma)]^{-2}$$

and

$$-2\sin^3\tfrac{1}{2}(\psi-\sigma)[\cos(\psi-\sigma)\cos\tfrac{1}{2}(\psi-\sigma)]^{-1}$$

respectively and therefore do not really diverge; moreover, they both vanish provided $\psi=\sigma$.

This seeming singularity is clearly caused by the chosen way of the approximating of the basic equations only. Indeed, if $\psi=\sigma$ the leading term of $\lambda_1$ expansion vanishes and the assumption $m/\lambda_1 \ll 1$ which we have used when estimated the integrals fails. The complete result remains however unaffected and no special care on that point seems to be necessary. Notice that a similar cancellation of the seeming singularities has been encountered in the reducing of the expression defining $\mathbb{B}$ as well.

Since the coefficient in the last term of (15) $(m\omega)\simeq 10^{-4}\epsilon$ is very small it is sufficient to keep only those parts of the whole $\mathbb{D}$ expression that significantly exceed the unit. Such terms exist only if $\sigma \ll 1$. The $\mathbf{B}_2$ aggregate contains the sum

$$-2\mathrm{cotan}\,\sigma-4\mathrm{cosec}\,2\sigma=-2(\cos\sigma+\sec\sigma)\mathrm{cosec}\,\sigma\approx -4/\sin\sigma.$$

Besides $\mathbb{D}$ contains itself the expression

$$-8\mathbf{A}_0\cos 2\sigma\ \ \mathrm{cosec}^2\,2\sigma-2\mathbf{A}_0\,\mathrm{cosec}^2\,\sigma,$$

but it is equal to

$$2[1-\frac{\sin^2(\chi/2)}{\sin^2\sigma}][2-\tan^2\sigma](\cos\sigma+\cos\chi/2)^{-1}$$

and does not considerably increases as $\sigma$ decreases (we would remind that $\sigma\geq\chi/2$). Thus the maximal contribution to $\mathbb{D}$ is due to the expression $-4\sin(\psi+\sigma)/\sin\sigma$ and one has

$$\mathbb{D}\approx -4\frac{\sin\psi}{\sin\sigma}, \tag{19}$$

provided $\sigma\ll 1$.

We see now that under the typical conditions even in the most favorable case

$$|(m\omega)\mathbb{D}|\leq 0.1\epsilon$$

and thus the contribution due to the last term in (15) is at least an order less than the current measurement uncertainty.

The formulae (15)–(19) yield the final expressions for the locally measurable deflection angle. In the case of free orbital motion of observer one has additionally $\omega r=\sqrt{m/r}$.

## 5. Discussion

Let us briefly discuss the separate contributions to the deflection angle (15). The sum

$$\delta_{\rm ab} = (\omega r)\,\mathbb{A} + (\omega r)^2\,\mathbb{C} \tag{20}$$

does not depend on the mass $m$ of central body and is determined mainly by the orbital speed $(\omega r)$ of the observer. It can be clearly classified as to describe the effect of the *relativistic aberration* calculated up to the second order of the speed (the third order contribution could be showed to be negligible). One may name it *kinematic* deflection angle as well. The following estimate holds for it in the case of the Sun-Earth system

$$0 < -\delta_{\rm ab} \leq 2\ 10^4\ \epsilon$$

The aberration contribution dominates in absolute value. However it is not involved as a rule in the theoretical analysis of the gravitational light deflection, perhaps, as a side effect existing even without any gravitational field. The origin of the aberrational contribution to the deflection angle is the different mutual orientations of the registered rays and the orbital speed of observer during the principal and the calibrating measurements respectively. Thus this part of the deflection angle is not really related to any light rays *deflection* and our terminology is perhaps not very adequate in this point. Anyway, the aberration dependent angle (20) contributes the experimentally measured difference of the 'undisturbed' and 'disturbed' arc separations between the master and reference sources and that is why it is kept in the deflection formula (15).

It is worthwhile to mention here that due to the Eq. (18a) the aberration contribution to the deflection angle formally depends on the *arc diameter* $\chi$ of the central body. This seems to be rather strange but really means only that if the measurements are carried out several times with *different* $\chi$ values but *constant* all the other conditions of experiment the corresponding values of $\sigma$ *will* slightly differ precisely in such a way to ensure the deflection angle (15) to remain unchanged. One may say that the direction *to the center* (described by $\sigma$) will depend on $\chi$ but there is no contradictions in such a statement because the 'irrelevant' parameter $\psi$ is not involved in this dependence.

The last term of (15)

$$\delta_{\rm g-k} = (m\omega)\mathbb{D} \tag{21}$$

is of a mixed *gravitational–kinematic* nature. For the Sun-Earth system its value does not exceed $0.1\epsilon$ and hence its detection should require at least 10-fold improvement of the measurement accuracy. At the recent level of the accuracy of experiments it can be temporarily omitted.

The most important result implied by the formula (15) is the expression for the most interesting *pure gravitational* contribution to the deflection angle which would coincide with the total deflection $\delta$ if there were no rotation of observer ($\omega = 0$) or, to be more exact, if his rotation were sufficiently slow:

$$\delta_{\rm gr} = \frac{2m}{r}(2\ \mathrm{cotan}\ \sigma)\frac{\sin\psi}{\sin\psi + \sin\sigma}. \tag{22}$$

It is useful to compare it with the classical expression (see [3], Eq. (A10), or [6], Eq. (40.11))

$$\tilde{\delta}_{\rm st} = \frac{2m}{r}\left(\mathrm{cotan}\frac{\sigma}{2}\right) \tag{23}$$

The main formal difference with (22) is the manner of entrance of the factor 2: *under* (in (23)) or *outside* (in (22)) of the *cotan* symbol. The 'compromise' between these two formulae is thus impossible as far as the function *tan* deflects from the linear proportionality.

To be more exact, the latter formula pretends to determine the *absolute* deflection angle, meanwhile we are operating with more physically meaningful *relative* one. Hence its differential version

$$\begin{aligned}\delta_{\rm st} &= \tilde{\delta}_{\rm st}(\sigma) - \tilde{\delta}_{\rm st}(\sigma+\zeta)\\ &= \tilde{\delta}_{\rm st}(\sigma) - \tilde{\delta}_{\rm st}(\sigma+\psi) + O(m\omega)\\ &\simeq \frac{2m}{r}[\mathrm{cotan}\frac{\sigma}{2} - \mathrm{cotan}\frac{\sigma+\psi}{2}]\\ &\equiv \frac{2m}{r}(\mathrm{cotan}\ \sigma + \mathrm{cosec}\ \sigma)\frac{2\sin\psi}{\sin\psi+\sin(\psi+\sigma)+\sin\sigma}\end{aligned} \tag{24}$$

is the more appropriate object for comparison.

The formulae (22) and (24) are clearly different though rather like. The leading terms of the expansion of their discrepancy with respect to $\sigma$ (assuming it to be small) are estimated as follows:

$$\delta_{\rm st} - \delta_{\rm gr} = \frac{2m}{r}[\tan\frac{\psi}{2} - \frac{\sigma}{2}\tan^2\frac{\psi}{2} + O(\sigma^2)] \approx 2\epsilon\tan(\psi/2). \tag{25}$$

and do *not* vanish even $\sigma$ is arbitrary small. Calculations with the complete formulae (22), (24) for $\psi = 60^\circ$ confirm that $\delta_{\rm st} - \delta_{\rm gr}$ is almost constant and slowly decreases from $0.29(4m/r)$ to $0.27(4m/r)$ (we would remind that accidentally $m/r \simeq \epsilon$) as $\sigma$ increases from 0.003 radians (source close to the edge of the limb) to $0.5\psi$. Thus it is of order of the measurement uncertainty under the typical conditions. It is also interesting to estimate the similar difference when the second order aberration term $(\omega r)^2\mathbb{C}$ and the cross gravitational–kinematic term $(m\omega)\mathbb{D}$ are taken into account as well (*i.e.* the complete deflection angle $\delta$ without the *first* order aberration term is used). The calculation shows that it increases approximately to $0.5\,(4m/r)$. Thus the difference $\delta_{\rm st} - \delta_{\rm gr}$ is close to the quadratic aberration contribution.

We shall not discuss here in the full details the origin of the defect of the classical formula (23) which is manifested in its disagreement with Eq. (22). We only remark that, as to our opinion, it has arisen due to the implicit attaching of too great importance to the Euclidean geometry of the sheet of paper where the theorists depicted the light rays and observer's trajectories.

More correctly, the connection between the space-time geometry and the results of observations has been built in the way inadequate to the effect under consideration. It seems nonsense to seek the mistakes in the (rather transparent) calculations in a number of wellknown textbooks and monographs, including refs. [6-13], analyzing the light deflection in the Schwarzschild geometry: these calculations are certainly correct. The discrepancy arises *before* the calculations are initiated. They are contained in the principles of description of the deflection angles measurement.

There is a lot of expositions of the calculations leading to the formula (23), but the analysis manifests the following their common feature: the interpretation in physical terms of the geometrical objects dealing with the inevitably *local* measurements of the light deflection involves some *nonlocal* facilities. As a rule the Minkowski space geometry is used although it is not sometimes explicitly declared. PPN method, which is used, for example, in [6,7,12-15], is directly based on the auxiliary Minkowski space. On the other hand in the book [8] the deflection angle is simply identified with the total variation $mod\ \pi$ of the coordinate $\phi$ along a null geodesic. This result can be equipped by an invariant geometrical meaning of course (this has been done in [5], for example) but it is clear that such an angle can be 'measured' only in terms of the geometry of the auxiliary flat space-time and is in no way a local quantity. [The nonlocal operation here is the parallel transport of the vector tangent to the ray from the beginning to the end of the ray in order to compare it with the vector tangent to the ray therein. The space through which transport is carried out must be flat, otherwise the result would depend on the path of transport.] Similar remark can be done with respect to refs. [9-11,16] as well.

Another typical example can be found in [17]. The straight lines constituting the angle $\varphi_0$ in the FIGURE 1 (p.69) do *not* exist in the curved space-time and one is forced to introduce the auxiliary Minkowski space in order to attach the geometrical meaning to the formal calculations therein. It cannot be realized in a unique way however.

It is worthwhile to note that similar pictures can be found in almost every work and they (*i.e.* the corresponding underlying flat space geometry) are really used in calculations.

And as the final example, in the pioneering paper [2] the working tool is the system of isotropic coordinates in the Schwarzschild space that are in fact interpreted as the polar coordinates in some *flat* space which also is not really manifested as the arena of a physical measurement.

We have therefore two weak points lying in the foundation of the Eq. (23):

(i) The nature of the geometrical tools used for its derivation is nonlocal contrary to the local nature of the observation that at least require a separate study, and

(ii) the Minkowski space cannot be uniquely distinguished in the curved space-time even if the curvature of space-time is small.

The latter fact has been mentioned in [18] but it has to be connected with inadequate approach based on "perturbations around flat space-time" rather than with the physical essence of the problem.

On the contrary, our approach does not reveal any such sort ambiguity. It is based in fact on the metric (3) alone, that simply realizes the principle of the constancy of the local light speed. In these frames, the introduction of auxiliary Minkowski space and further development of a perturbation scheme are unnecessary (and inadequate) complications in fact *.

---

* This is a reason why it seems to be not very fruitful to perform a detailed comparison of our aberration formula (Eqs. (16),(18),(20)) with its standard counterpart which can be found, for example, in [12], p 64, because the latter one is also based on the PPN approach.

Some uncertainty that exists due to possibility of different definitions of what moment of time is most reasonable for calibrating measurement (see Section 2) is connected with the physical essence of problem and can be removed, for example, by a reasonable convention.

Additionally to the Eq. (22), the gravitational deflection angle for 'bisectorial' choice of the moment of calibrating measurement is calculated in the Appendix (Eq. (A22)). It also does not coincide with the standard formula (24), the discrepancy being of order $\epsilon(4\tan\frac{1}{4}\psi)$, *i.e.* similar to that in Eq. (25).

Finally, it must be emphasized that the validity of our approach is in no way related to the above speculations concerning with the other more usual approaches and must be estimated independently on their estimate.

## 6. Resume

In this work we have found the basic equations determining the value of the angle of light deflection as it is registered by the observer uniformly rotating in a static spherical symmetric gravitational field along a circular orbit (Eqs. (6-8)). Their derivation is performed in frames of the standard general relativity and is based exclusively on the principle of the constancy of the local speed of light. In particular it does not involve any artificially constructed 'frames of reference', Fermi-Walker transport, PPN-approximations or similar tools.

The result has been further fitted to the case of measurement in the Earth-Sun system from the Earth's orbit and the formulae sufficiently simple for applications have been obtained (Eqs. (15-18)).

The whole deflection angle is divided into the several contributions: the contribution of the relativistic aberration (20) (kinematic deflection of the first and second orders), the pure gravitational deflection (22) and the mixed gravitational-kinematic deflection (21).

Our calculations prove that the classical formula $\tilde{\delta}_{st} = (2m/r)\text{cotan}(\sigma/2)$ does not determine the gravitational deflection comprehensively even if it is adopted to a really used differential measurement scheme and the rotation of observer can be neglected.

The necessary correction to the gravitational deflection is expected to be of order $\epsilon$ where $\epsilon \simeq 10^{-8}$ (radians) is the level of the uncertainty for the recent radio-waves deflection measurements by means of the VLBI method. Thus one may hope that the this discrepancy can be detected by means of the up-to-day or perhaps somewhat improved experimental technic.

This task seems to put forward rather complicated problems however. At first, there is a number of masking effects affecting the observed deflection angle that we have not touched on in this paper at all (for example, the gravitational field and the own rotation of the Earth, the nonzero eccentricity of the Earth's orbit, the refraction of the atmosphere of the Earth and the solar corona etc., see also [4]). They are to be taken in account independently if one intends to analyze the real experimental data. The crucial point, however, seems to be a necessity of a sufficiently accurate estimation of the angle $\sigma$, *i.e.* the observed direction to the center of the Sun in fact. The VLBI method seems to be not very useful for this purpose and perhaps some new ideas should be drawn in.

Finally it is worthwhile to notice that our requirement for the both observed sources to belong to the ecliptic could be easily removed resulting some geometry complication only. This should somewhat improve the model plausibility.

**Acknowledgment**. I am grateful to the *Graduertenkolleg* “Scientific computing” (Cologne – St. Augustin, the North-Rhein-Vestfalia) for financial support and to the Institute of Theoretical Physics of the University of Cologne for hospitality.

## Appendix

Here we calculate the light deflection angle in the case of alternative choice of the moment of ‘minimally disturbed’ positions of the master and reference sources on the celestial sphere which has been mentioned in the Section 2. Now we shall assume that the bisector of the angle between the sources is precisely of the outgoing radial direction. The scheme of observations is depicted in the Fig. 2. Its only difference with the Fig. 1 is the position of the point of calibrating observation which will be now denoted **C’**. It does not lie generally speaking on the ray $\mathcal{L}$ but rather on another ray $\mathcal{L}$’ from the master source. Obviously the position of **C’** is uniquely determined by the congruences of rays from the sources and does not depend on the choice of the moment of principal observation **P**.

The calculations of the light deflection angle in the case **C'** reproduce *mutatis mutandis* ones in the Sections 2 and 3.

Geometrically, the point **C** is replaced by the point **C'**, the geodesic which corresponds to the ray $\mathcal{L}$' being characterized by the $\lambda$-parameter $\lambda_1$ instead of $\lambda_0$ for $\mathcal{L}$.

Analytically, the first equation of the set (6) is replaced by the equation

$$\psi = \arctan U(\lambda_1, -\omega) + \arctan U(\lambda_1, \omega) \tag{A6}$$

meanwhile other equations (6) still hold. (We use the notations borrowed from the Sections 2,3.)

Instead of Eq. (8) one has now

$$2I(r,\lambda_1) + I(r,\lambda_0) - I(r,\lambda_2) + 2[-I(R(\lambda_0),\lambda_0) + I(R(\lambda_2),\lambda_2)] = 0 \tag{A8}$$

that can be reduced in the first approximation with respect to $m/r$ to

$$\begin{aligned} 2\alpha_1 + \alpha_0 - \alpha_2 + \frac{m}{r}\{ \; & 2(2\mathrm{cosec}\,\alpha_1 - \mathrm{cosec}\,\alpha_0 + \mathrm{cosec}\,\alpha_2) \\ & -2(2\mathrm{cosec}\,2\alpha_1 + \mathrm{cosec}\,2\alpha_0 - \mathrm{cosec}\,2\alpha_2) \\ & -(2\mathrm{cotan}\,\alpha_1 + \mathrm{cotan}\,\alpha_0 - \mathrm{cotan}\,\alpha_2)\} \simeq 0. \end{aligned} \tag{A11}$$

All the other relevant equations will follow from ones mentioned above.

Here however we shall be interested only in the *pure gravitational* contribution to the deflection angle and hence assume the rotation of observer to be sufficiently slow to neglect it everywhere. Additionally, this enable us to avoid the necessity of the special definition of the 'radial direction'. We omit therefore all the terms involving the rotation frequency parameter $\omega$.

The deflection angle is determined by the equation

$$\delta_{\max} = \psi - \zeta \simeq 2\alpha_1 + \alpha_0 - \alpha_2 - \frac{m}{r}(2\tan\alpha_1 + \tan\alpha_0 - \tan\alpha_2), \tag{A9}$$

which in view of ($A11$) takes the form

$$\begin{aligned} \delta_{\max} = \frac{m}{r}\{ & -2(2\mathrm{cosec}\,\alpha_1 - \mathrm{cosec}\,\alpha_0 + \mathrm{cosec}\,\alpha_2) \\ & -(2\tan\alpha_1 + \tan\alpha_0 - \tan\alpha_2) \\ & \; 2(2\mathrm{cosec}\,2\alpha_1 + \mathrm{cosec}\,2\alpha_0 - \mathrm{cosec}\,2\alpha_2) \\ & +(2\mathrm{cotan}\,\alpha_1 + \mathrm{cotan}\,\alpha_0 - \mathrm{cotan}\,\alpha_2)\}. \end{aligned} \tag{A12}$$

It remains to express $\alpha_j$ in terms of angles $\sigma, \psi$, *i.e.* to derive the counterpart of Eqs. (14). It is now sufficient however to derive their *zero order* approximation. In the same way as in the Section 3 one easily obtains

$$\begin{aligned} \alpha_0 &\simeq \sigma, \\ \alpha_1 &\simeq \tfrac{1}{2}\psi, \\ \alpha_2 &\simeq \sigma + \psi \end{aligned} \tag{A14}$$

up to the terms $O(m/r)$ (that would yield $O((m/r)^2)$ contribution to $\delta_{\rm max}$ which is neglected).

After the substitution for $(A14)$ to $(A12)$ and some transformations one obtains

$$\begin{aligned}\delta_{\rm max} &\simeq \delta_{\rm st} - \frac{4m}{r}\tan\tfrac{1}{4}\psi \\ &\equiv \frac{2m}{r}{\rm cosec}\,\tfrac{1}{2}\sigma\ \frac{2\tan\tfrac{1}{4}\psi}{\sin\tfrac{1}{2}(\psi+\sigma)}\cos^2\tfrac{1}{4}(\psi+2\sigma)\end{aligned} \qquad (A22)$$

that is the desirable result.